\documentclass[epj]{webofc}
\usepackage[utf8]{inputenc}
\usepackage[varg]{txfonts}   % Web of Conferences font
\usepackage{booktabs}
\usepackage{xcolor}
\definecolor{darkred}{rgb}{0.4,0.0,0.0}
\definecolor{darkgreen}{rgb}{0.0,0.4,0.0}
\definecolor{darkblue}{rgb}{0.0,0.0,0.4}
\usepackage[bookmarks,linktocpage,colorlinks,
    linkcolor = darkred,
    urlcolor  = darkblue,
    citecolor = darkgreen]{hyperref}

\usepackage{epsfig}
\usepackage{geometry}

\usepackage{subfigure}

\wocname{EPJ Web of Conferences}
\woctitle{Lattice2017}

\def\msbar{\overline{\rm MS\kern-0.5pt}\kern0.5pt}
\def\lmatrix{\left(\begin{array}}
\def\rmatrix{\end{array}\right)}
\def\bea{\begin{eqnarray}}
\def\eea{\end{eqnarray}}
\def\nn{\nonumber}

\begin{document}

\selectlanguage{english}

\title{
Weakly coupled conformal gauge theories on the lattice
}

\author{
\firstname{Zoltan} \lastname{Fodor}\inst{1,2,3} \and
\firstname{Kieran} \lastname{Holland}\inst{4} \and
\firstname{Julius} \lastname{Kuti}\inst{5} \and
\firstname{Daniel} \lastname{Nogradi}\inst{3,6,7}\fnsep\thanks{Speaker, \email{nogradi@bodri.elte.hu}} \and
\firstname{Chik Him} \lastname{Wong}\inst{1} 
}

\institute{
University of Wuppertal, Department of Physics, Wuppertal D-42097, Germany
\and
Juelich Supercomputing Center, Forschungszentrum Juelich, Juelich D-52425, Germany
\and
Eotvos University, Pazmany Peter setany 1/a, 1117 Budapest, Hungary
\and
University of the Pacific, 3601 Pacific Ave, Stockton CA 95211, USA
\and
University of California, San Diego, 9500 Gilman Drive, La Jolla, CA 92093, USA
\and
MTA-ELTE Lendulet Lattice Gauge Theory Research Group, 1117 Budapest, Hungary
\and
Universidad Autonoma, IFT UAM/CSIC and 
Departamento de Fisica Teorica, 28049 Madrid, Spain
}

\abstract{
Results are reported for the $\beta$-function of weakly coupled conformal gauge theories on the lattice, SU(3)
with $N_f = 14$ fundamental and $N_f = 3$ sextet fermions. The models are
chosen to be close to the upper end of the conformal window where perturbation theory is reliable hence a fixed point is
expected. The study serves as a test of how well lattice methods perform in the weakly coupled conformal cases.
We also comment on the 5-loop $\beta$-function of two models close to the lower end of the conformal window, 
SU(3) with $N_f = 12$ fundamental and $N_f = 2$ sextet fermions.
}

\maketitle

\section{Introduction}
\label{intro}

We study gauge theories inside the conformal window close to its upper end. In this region the gauge group, number of
massless fermion flavors and the representation they carry are such that the perturbative $\beta$-function possesses a
fixed point at a small value of the renormalized coupling. Hence it is expected that the loop expansion is reliable and
the existence of the infrared fixed point will not be spoiled by any non-perturbative effect. The corresponding models
are genuine interacting conformal field theories with 
non-trivial (small) anomalous dimensions \cite{Caswell:1974gg, Banks:1981nn}.

Our main motivation is that there has been extensive lattice study of models close to the lower end of the conformal
window because of their relevance for BSM model building in recent years \cite{Nogradi:2016qek}. 
Close to the lower end of the conformal window
non-perturbative effects are relevant because in the conformal case the fixed point coupling is large and in the chirally
broken case the entire low energy dynamics is dictated by non-perturbative effects similarly to QCD. In principle
lattice simulations are an ideal tool to determine whether a given model is inside or outside the conformal window
exactly because the lattice setup can capture all non-perturbative effects. 
Nevertheless systematic effects can be large sometimes leading to
controversies for models close to the lower end
of the conformal window. 
We study here the weakly coupled conformal case which can be
tested on the lattice with predictable and controlled results in sufficient
orders of perturbation theory.
We are interested in how the lattice tool set
is able to identify conformality, if there are any unexpected systematic effects and how ambiguous or unambiguous the
lattice results are.

In this work the finite volume gradient flow \cite{Luscher:2010iy} running coupling 
scheme \cite{Fodor:2012td, Fodor:2012qh} is used 
to calculate the $\beta$-function and to probe the infrared dynamics.

\section{Two weakly coupled conformal theories}
\label{twomodels}

\begin{figure}
\centering
\includegraphics[width=12cm]{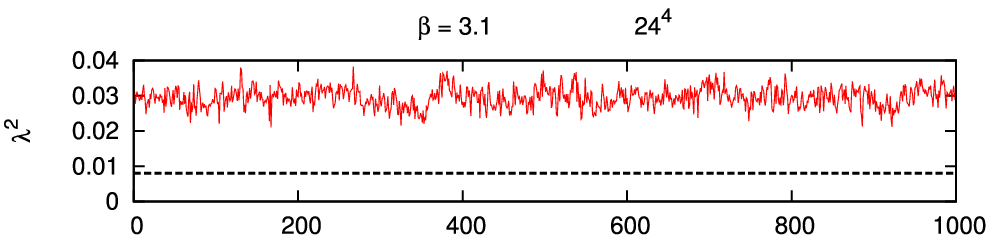} \includegraphics[width=12cm]{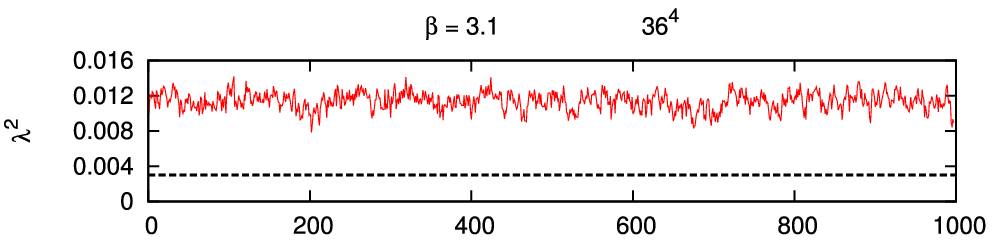}
\caption{Monte Carlo history of the lowest Dirac operator, $D^\dagger D$, eigenvalues for the $N_f = 14$ fundamental
model at the lowest bare coupling
$\beta$ on two selected volumes. Every $10^{th}$ configuration was measured, the runs correspond
to 10000 trajectories in total. The dashed lines show the chosen lower bound 
for the Remez algorithm. The histories for different bare coupling $\beta$ and the $N_f = 3$ sextet model are similar.}
\label{eigs}
\end{figure}

The models we study are both SU(3) gauge theories. One of them has $N_f = 14$ fundamental, the other
$N_f = 3$ sextet (two-index-symmetric) fermions. In what follows we will 
sometimes refer to the first simply as $N_f = 14$ and the latter as
$N_f = 3$, without specifying the representation. The $\beta$-functions in $\msbar$ are known to 5-loops
\cite{Baikov:2016tgj, Herzog:2017ohr},
\bea
\mu^2 \frac{dg^2}{d\mu^2}  = \sum_{i=1}^5 b_i \frac{g^{2i+2}}{(16\pi^2)^i} 
\eea
where the coefficients $b_i$ for the two models are
\bea
N_f = 14:&&\quad b_1 = -\frac{5}{3},\;\; b_2 = \frac{226}{3},\;\;  b_3 = \frac{70547}{54},\;\;\; b_4 = -15506.48,\;\;
b_5 = -668754.5 \nn \\
N_f = 3:&&\quad b_1 = -1,\;\;\; b_2 = 148,\;\;\; b_3 = \frac{3493}{2},\;\;\;\;\, b_4 = -22834.07,\;\; b_5 = -2365262.5  \nn
\eea
The corresponding fixed points at increasing loop order are then simply
\bea
N_f = 14:&&\quad g_{*2}^2 = 3.494,\quad g_{*3}^2 = 2.696,\quad g_{*4}^2 = 2.810,\quad g_{*5}^2 = 2.926 \nn \\
N_f = 3:&& \quad g_{*2}^2 = 1.067,\quad g_{*3}^2 = 0.993,\quad g_{*4}^2 = 0.999,\quad g_{*5}^2 = 1.002 
\eea
Clearly, these fixed points are all rather small and do not change much beyond 3-loops so it is reasonable to expect
that the loop expansion is already a good approximation. This is especially the case for the $N_f=3$ sextet model, which is closer
to the upper end of its conformal window than the $N_f=14$ fundamental model. For the sextet model asymptotic freedom is lost
at $N_f = 3.3$ while for the fundamental model at $N_f = 16.5$. Actually $N_f = 14$ was chosen so that it is further from the
upper end but still such that the perturbative calculation is presumably trustworthy. Not being very close to the upper
end of the conformal window will have practical consequences as explained in the next section. 

The fact that the fixed points in both models are expected to be at small couplings in $\msbar$ indicates that
scheme dependence is also expected to be small. Hence a comparison with our scheme (which is different from $\msbar$)
is meaningful.

\begin{figure}
\centering
\includegraphics[width=4.3cm]{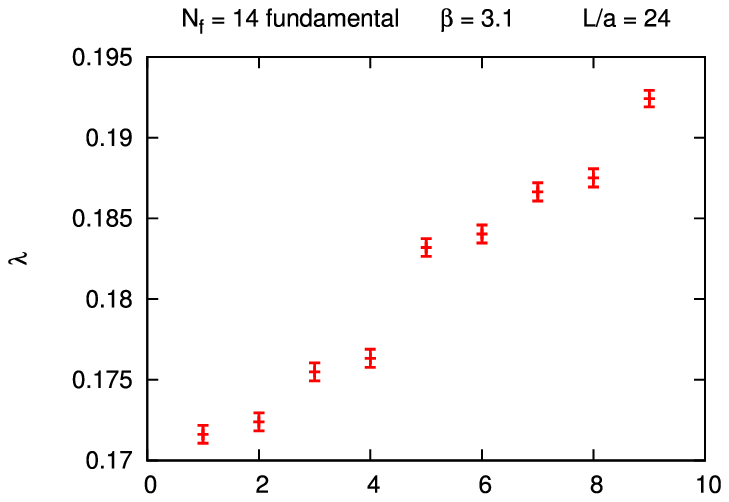} \includegraphics[width=4.3cm]{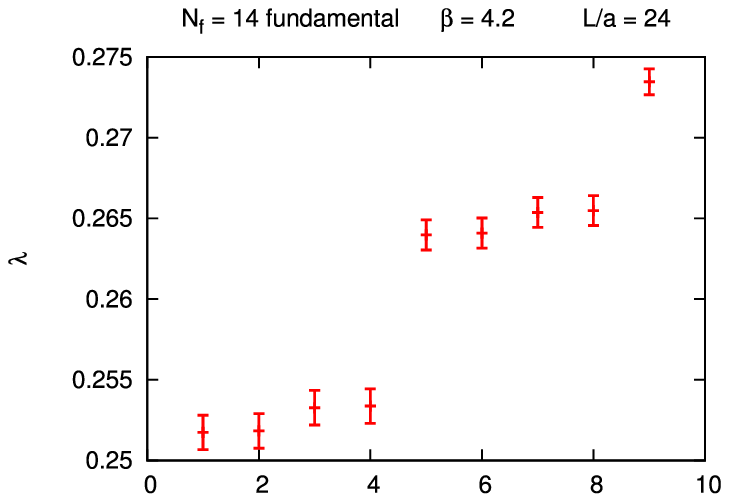} \includegraphics[width=4.3cm]{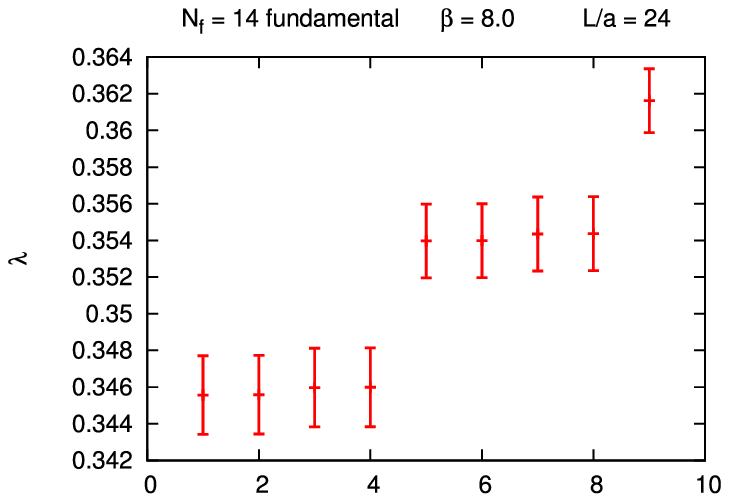}
\includegraphics[width=4.3cm]{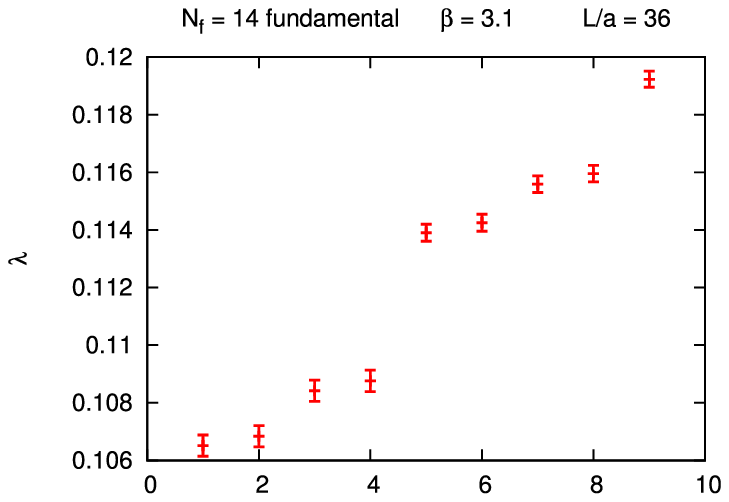} \includegraphics[width=4.3cm]{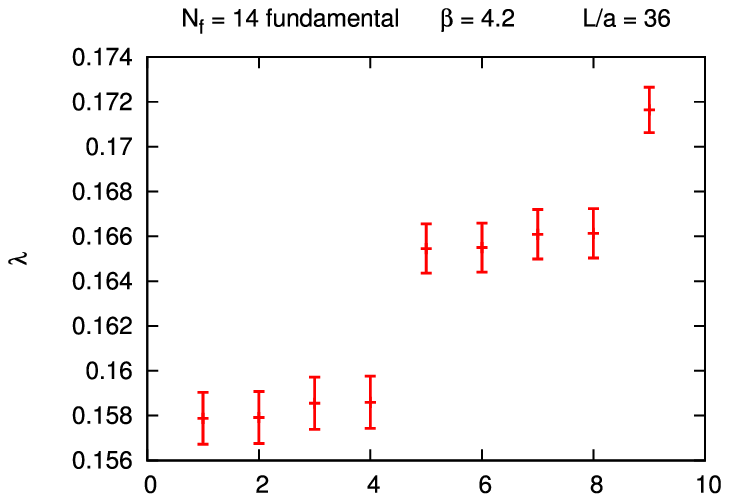} \includegraphics[width=4.3cm]{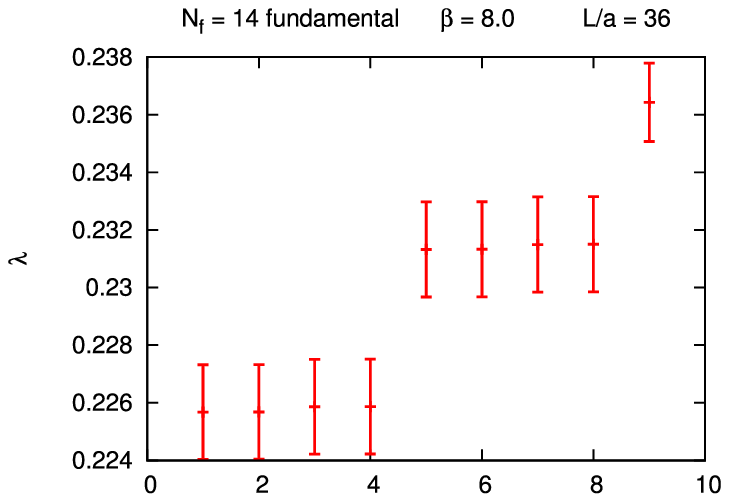}

\includegraphics[width=4.3cm]{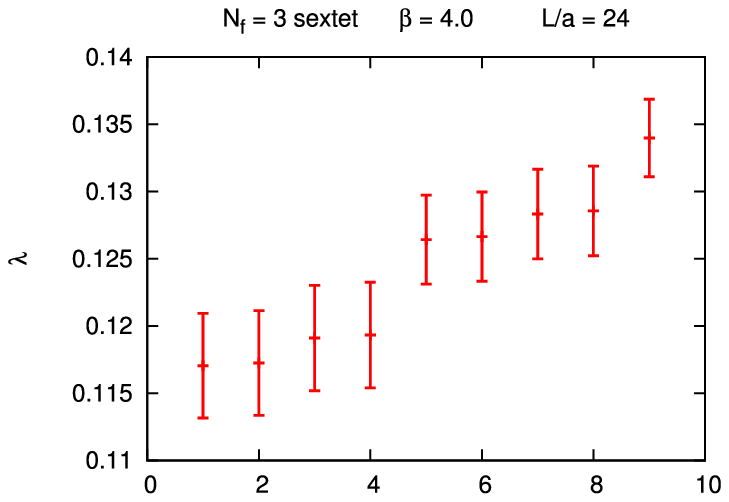} \includegraphics[width=4.3cm]{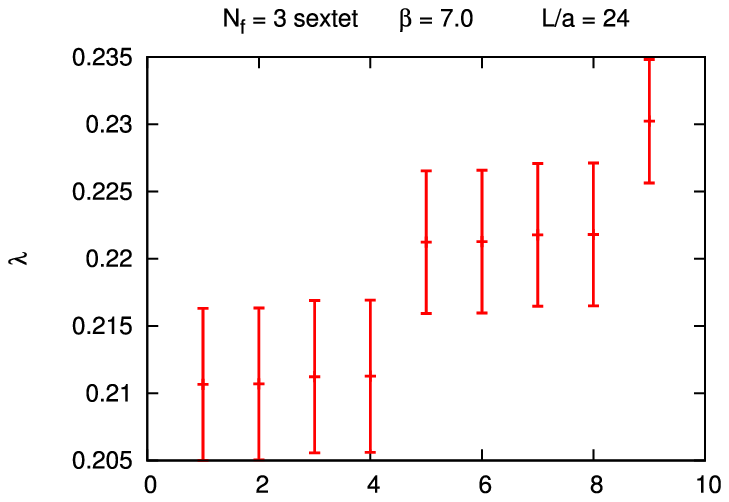} \includegraphics[width=4.3cm]{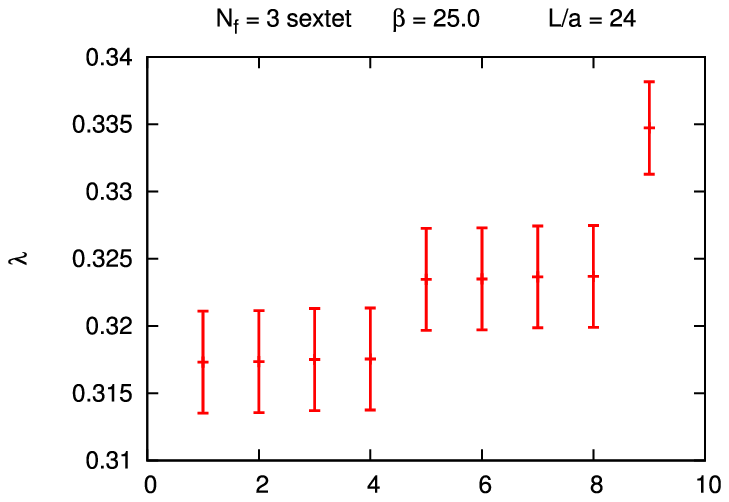}
\includegraphics[width=4.3cm]{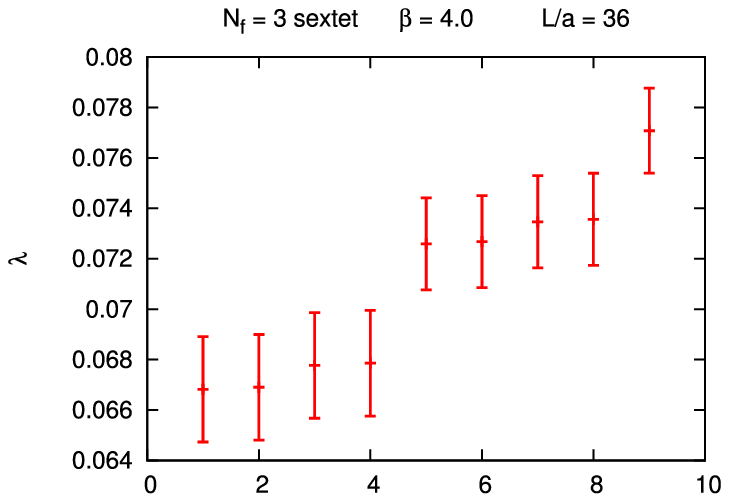} \includegraphics[width=4.3cm]{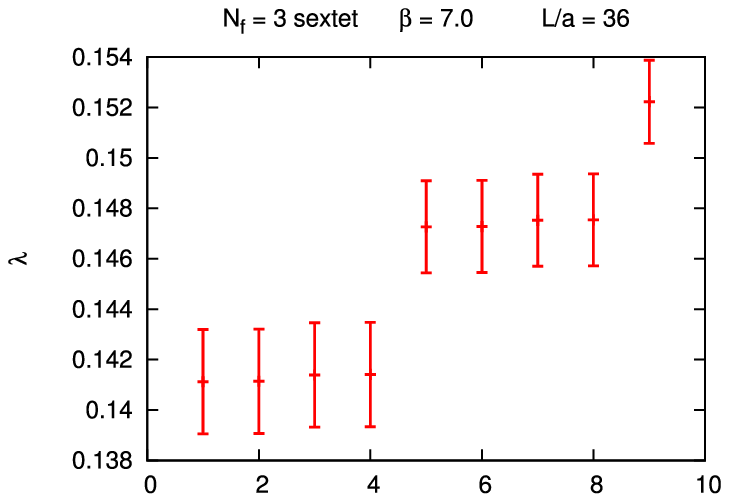} \includegraphics[width=4.3cm]{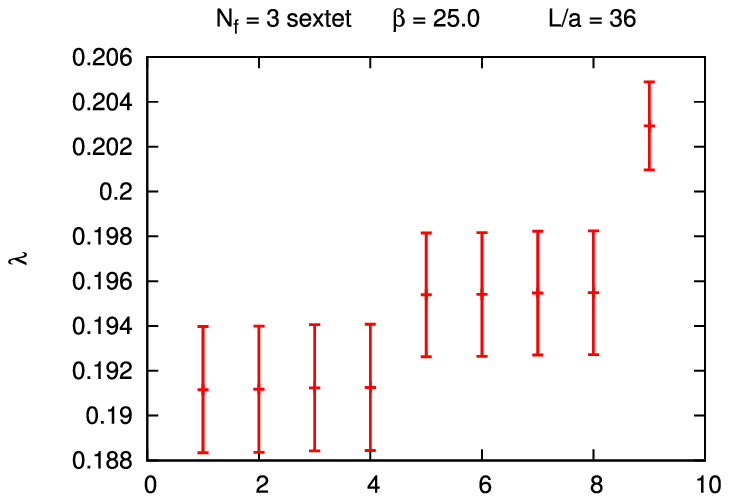}
\caption{The lowest 9 eigenvalues for both models at the lowest, medium and largest bare coupling $\beta$ on two selected
volumes. Top two rows: $N_f = 14$ fundamental, bottom two rows: $N_f = 3$ sextet. Bare coupling $\beta$ is increasing from left
to right.}
\label{quartets}
\end{figure}

\section{Numerical simulation}
\label{numericalsimulation}

The simulations are carried out via the staggered discretization using stout improvement. The running coupling is
defined in the finite volume gradient flow scheme with periodic gauge fields and fermions which are anti-periodic in all
four directions. The coupling in this scheme is given by
\bea
g^2(L) = \frac{128\pi^2}{3(N^2-1)\left(1+\delta(c)\right)} \langle t^2 E(t) \rangle
\label{g2}
\eea
where $N=3$, $c=\sqrt{8t}/{L}$ is a constant, $\delta(c)$ is a known factor and we set $c=1/5$ for 
definiteness; see \cite{Fodor:2012td, Fodor:2012qh} for more details.
Due to finite volume and the remnant chiral symmetry of staggered fermions the bare mass can be set to zero, $m=0$,
and the only parameters are the lattice volume and the bare coupling $\beta$.

Since the flavor content in neither model is divisible by 4 the rooting procedure is required, implemented by the RHMC
algorithm. A necessary ingredient is the Remez algorithm which requires preset values for the interval on which the
fourth root will be approximated by a rational function. Since the mass is zero, a good choice for the lower end of 
this interval needs to be measured first. This is shown in figure \ref{eigs} for two examples. Once an appropriate 
lower bound is found the simulation is stable. The validity of the rooting procedure at zero fermion mass and finite volume
was described in detail in \cite{Fodor:2015zna}. Key is the finite gap in the Dirac spectrum due to the anti-periodic
boundary conditions of the fermions. As the continuum is approached taste breaking is reduced and already at the lowest
bare coupling $\beta$ the quartets in the Dirac spectrum are clearly visible. We measured the lowest 9
eigenvalues and show examples of the restoration of quartet degeneracy in figure \ref{quartets}.

\begin{figure}
\centering
\includegraphics[width=7cm]{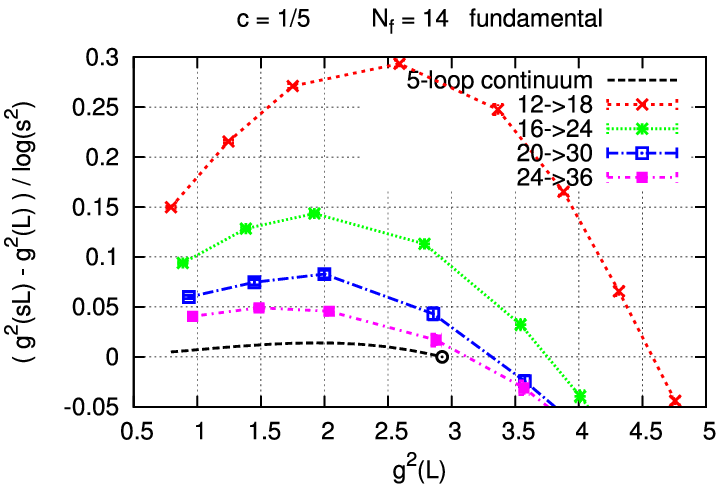} \includegraphics[width=7cm]{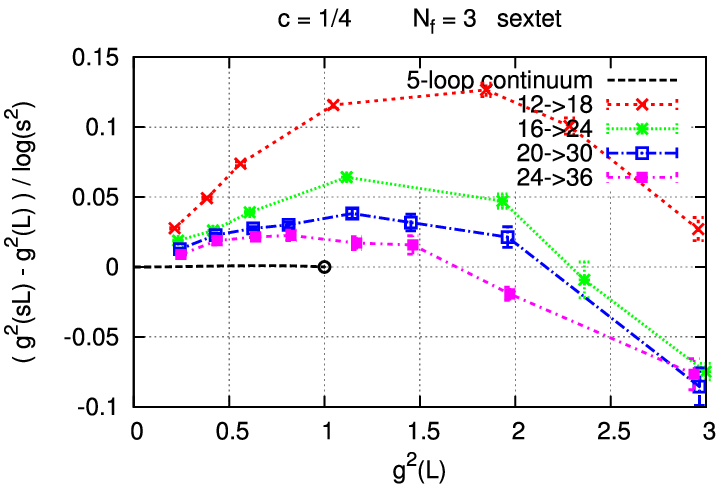}
\caption{Measured discrete $\beta$-function as a function of $g^2(L)$ for the $N_f = 14$ fundamental (left)
and the $N_f = 3$ sextet model (right).
The 5-loop $\msbar$ continuum result is also shown. Note that the maximum of the latter is quite small,
$\simeq 0.014$ (left) and $\simeq 0.001$ (right) hence hardly visible on the plot.}
\label{nf14betanf3beta}
\end{figure}

The discretization of the observable $E$ in equation (\ref{g2}) is done by the symmetric clover definition while the Symanzik tree
level improved gauge action is used both for the dynamical gauge action in the simulations and for the evolution of the
gradient flow. In the terminology of \cite{Fodor:2014cxa, Fodor:2014cpa} this setup corresponds to the SSC discretization.

Once the renormalized couplings are measured the discrete $\beta$-function, $( g^2(sL) - g^2(L) ) / \log( s^2 )$ 
for some finite ratio $s$ is straightforward to obtain. 
We choose $s=3/2$. The continuum will be approached by four pairs of lattice volumes, $12 \to 18$, $16 \to 24$, $20
\to 30$ and $24 \to 36$. The renormalized coupling is measured at various bare couplings on all volume pairs and the
results are shown in figure \ref{nf14betanf3beta}. Also shown in the plot is the
5-loop $\msbar$ result for comparison. Note that the maximum of the $\beta$-function in the $\msbar$ 5-loop case is
rather small, around $0.014$ and $0.001$ for the $N_f = 14$ and $N_f = 3$ models, respectively. The maximum being very
small in the latter is a direct consequence of its closeness to the upper end of the conformal window. Also note that 
the 5-loop beta-function possesses a second zero too, but at relatively
large coupling for both models, $g^2 \simeq 7.32$ for $N_f=14$ and $g^2 \simeq 6.39$ for $N_f=3$.
More importantly the location of the second zero at 4-loops is at $g^2 \simeq 18.6$
and $19.8$, respectively, hence convergence is not reached at the 5-loop level for
these second zeros, unlike for the first zero. Non-perturbative lattice calculations are needed to rule in
or rule out the second zero in the $\beta$-function as hinted from 5-loops at
strong coupling.

\begin{figure}
\centering
\includegraphics[width=10cm]{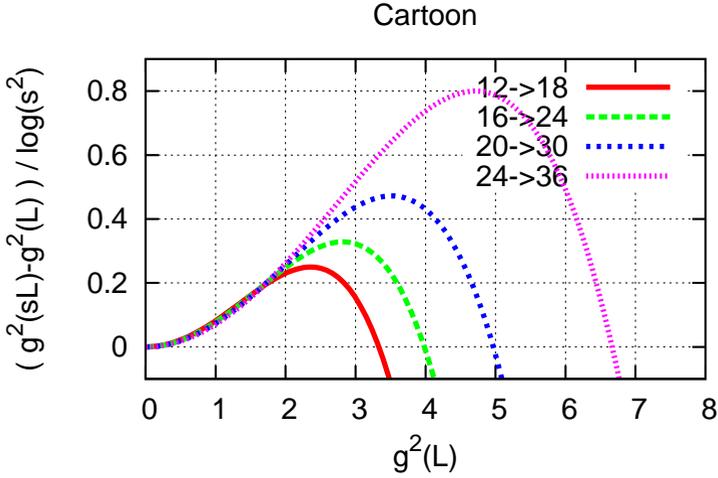}
\caption{A sketch (not real data) illustrating cases where the zero of the $\beta$-function at finite lattice volume
keeps increasing towards the continuum. In these cases it is a very delicate issue whether the zeros converge to a
finite value or run off to infinity. In the latter case the continuum theory is not conformal.}
\label{cartoon}
\end{figure}

A number of observations are in order. Both models exhibit the following property: the discrete $\beta$-function
possesses a zero on finite lattice volumes. One needs to be very careful about its interpretation though because it
is entirely possible that towards the continuum limit these zeros disappear. In particular if the location of the zeros is
{\em increasing} as the lattice volume is increasing, i.e. towards the continuum limit, then it is not at all clear whether
the zeros converge somewhere finite or run off to infinity. Hence a fully controlled continuum extrapolation is
mandatory in these cases before definite conclusions can be drawn. A sketch is shown in figure \ref{cartoon}
illustrating the potential problem.

Note further though that in both models the zeros of the $\beta$-function at finite lattice volume are such that they
are {\em decreasing} as the lattice volume is increasing. In other words, closer and closer to the continuum the
detected zero becomes smaller and smaller. In this case, contrary to the situation sketched in figure \ref{cartoon}, it
is rather easy to conclude assuming that the observed trend does not change. This is because in any case we know that
both models are asymptotically free hence a positive $\beta$-function is guaranteed at some small positive $g^2$. Hence
the convergence of the zeros to a positive and finite $g_*^2$ is guaranteed as $L/a \to \infty$, i.e. the continuum
model is conformal.

Finally, it must be emphasized that an important assumption is absolutely necessary for the above argument, namely that
the observed trend of decreasing zeros with increasing lattice volume does not change as the continuum is approached
further. If this assumption turns out to be wrong then the conclusion about conformality was premature. 

The actual continuum limit of the $\beta$-function at fixed $g^2(L)$ as performed usually is extremely demanding given
our data. This is because as noted above the expected maximum of the $\beta$-function is tiny and one needs to resolve
this tiny value from zero, in order to establish a positive continuum result. This requires very small absolute errors
on the renormalized couplings. Our current errors even though rather small in relative terms, are far too large in
absolute value for this. Hence we are not able to perform a controlled continuum extrapolation that is precise enough. 

\begin{figure}
\centering
\includegraphics[width=10cm]{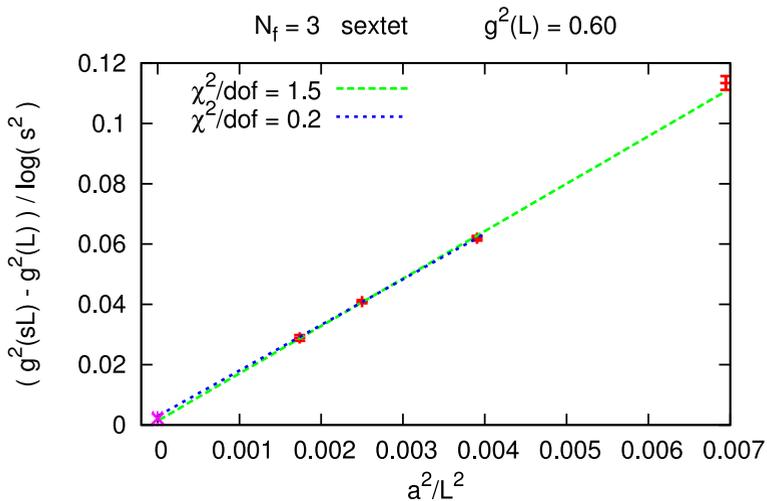}
\caption{Continuum extrapolation of the discrete $\beta$-function in the $N_f = 3$ sextet model after interpolating to
$g^2(L) = 0.60$. Two types of extrapolations were performed, either all 4 lattice spacings are used or the roughest one
is dropped. The difference is taken into account as a systematic error. 
The $\chi^2/dof$ of the extrapolations are shown in the legend.}
\label{cont}
\end{figure}

For instance, using our data we can interpolate the discrete $\beta$-function from nearby data points 
to $g^2(L) = 0.6$ on all lattice volume pairs for the $N_f = 3$ sextet model using various polynomial orders. The
variation of the interpolated values with the polynomial order is taken into account as a systematic error
and is added to the statistical error in quadrature. The obtained discrete $\beta$-function values are then extrapolated
to the continuum linearly in $a^2/L^2$. 
We may take the continuum limit using all 4 data points or by
dropping the roughest one and using only 3. 
Both have very good $\chi^2/dof$ and hence the difference in the final result
is again taken as a systematic error which is added to the statistical one; see figure \ref{cont}. 
We obtain $0.002(2)$ in this particular
example for the continuum value 
which is clearly consistent with the expected $0.001$ from the 5-loop $\msbar$ result but is also consistent
with zero hence is not very predictive. At other values of $g^2(L)$ our observations are similar.

Unfortunately this
property, namely that the maximum of the $\beta$-function is very small, 
seems to be a common feature of all weakly coupled gauge theories as it is a direct consequence of being close
to the upper end of the conformal window.

\section{Moving away from the weakly coupled regime}
\label{5loop}

Now that the result for the 5-loop $\beta$-function in $\msbar$ is available \cite{Baikov:2016tgj, Herzog:2017ohr}
it is enlightening to see what happens to
the perturbative predictions as the flavor number is decreased towards the lower end of the conformal window. We have
seen that the $N_f = 14$ fundamental and $N_f = 3$ sextet models are conformal with a perturbatively accessible fixed
point. Here we will show the perturbative behavior of the $N_f = 12$ fundamental and $N_f = 2$ sextet models, both being
actively studied on the lattice \cite{Hansen:2017ejh,Hansen:2016sxp,Drach:2015sua, Fodor:2016pls, Fodor:2016wal,
Fodor:2015zna, Cheng:2014jba, Hasenfratz:2016dou, Hasenfratz:2017mdh, Fodor:2016zil, Fodor:2017gtj}.

\begin{figure}
\centering
\includegraphics[width=6cm]{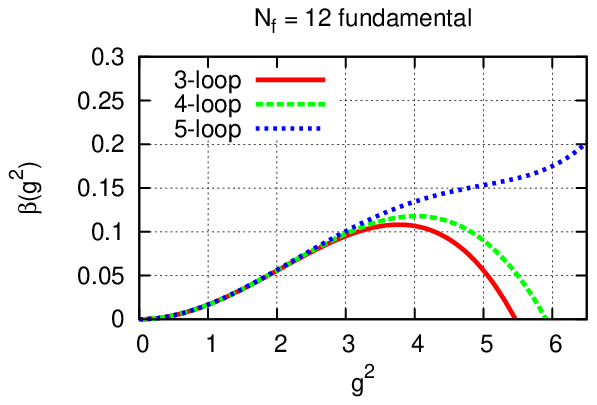}  \includegraphics[width=6cm]{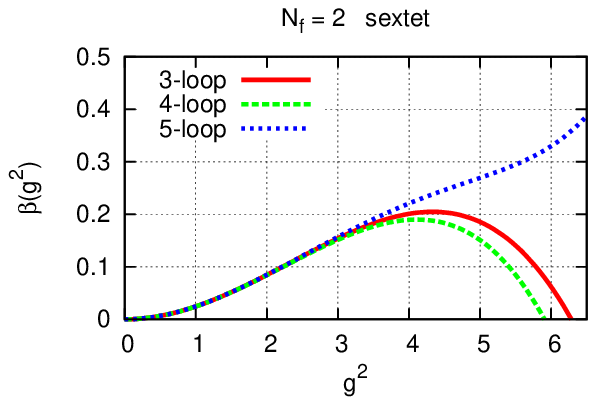}
\caption{The $\beta$-function of the $N_f = 12$ fundamental model (left) and the $N_f = 2$ sextet model (right) 
in $\msbar$ for increasing loop orders.}
\label{nf12nf2s}
\end{figure}

The perturbative $\beta$-function is shown in figure \ref{nf12nf2s} for the two models. They both share the
property that the 2-loop, 3-loop and 4-loop $\beta$-functions all have zeros, but at relatively large coupling, $g_*^2 >
5$. The 5-loop $\beta$-function is however without a fixed point suggesting QCD-like behavior, at least to this order of
the loop expansion. 

It is also straightforward to obtain the lower end of the conformal window at 5-loops order 
using the results \cite{Baikov:2016tgj, Herzog:2017ohr}. One obtains $N_f \simeq 12.89$ and $N_f \simeq 2.35$,
for the fundamental and sextet representation, respectively. How trustworthy these perturbative predictions are can of
course only be determined once all systematic effects are fully controlled in non-perturbative lattice simulations.

\section{Conclusion}
\label{conclusion}

Studies of gauge theories close to the lower end of conformal window are plaqued by large systematic effects. 
The difficulty of studying models far away from the weakly coupled CFT regime manifests itself both in the continuum via
the loop expansion and non-perturbative simulations. In lattice studies the zeros of the
$\beta$-function at finite lattice volumes may disappear towards the continuum signaling a qualitative change.
Similarly, the zeros of the continuum $\beta$-function may disappear with increasing loop order again signaling a
qualitative change in the supposed infrared dynamics. Hence it is very important to not completely trust the first few
loop orders
in the continuum calculation and also to go beyond small or medium size lattice volumes in lattice calculations.

In the weakly coupled CFT regime however the perturbative results for the $\beta$-function and its zero 
show little sensitivity to the order of the loop expansion. On the lattice the continuum extrapolation from finite
lattice volumes is still challenging because the expected $\beta$-function is small and hence all measured errors must
be small in absolute terms (not just relative) in order to resolve the result from zero. Nevertheless in our study of
two CFT cases, $N_f = 14$ fundamental and $N_f = 3$ sextet we have identified trends which if they do not change further
towards the continuum then guarantee the existence of a fixed point in the continuum. 

Even though the continuum limit of the $\beta$-function is challenging it may very well be that the mass anomalous
dimension $\gamma_*$ can be obtained more reliably. We hope to return to this question in a future publication.

\section*{Acknowledgement}

We acknowledge support by the DOE under grant DESC0009919,
by the NSF under grants 1620845 and
1318220, by OTKA under the grant OTKA-NF-104034,
and by the Deutsche Forschungsgemeinschaft grant SFBTR
55. Computational resources
were provided by the DOE INCITE program on the
ALCF BG/Q platform, USQCD at Fermilab, by the University
of Wuppertal, by Juelich Supercomputing Center on Juqueen
and by the Institute for Theoretical 
Physics, Eotvos University.


\begin{thebibliography}{99}

\footnotesize

%\cite{Caswell:1974gg}
\bibitem{Caswell:1974gg} 
  W.~E.~Caswell,
  %``Asymptotic Behavior of Nonabelian Gauge Theories to Two Loop Order,''
  Phys.\ Rev.\ Lett.\  {\bf 33}, 244 (1974).
  % doi:10.1103/PhysRevLett.33.244
  %%CITATION = doi:10.1103/PhysRevLett.33.244;%%
  %890 citations counted in INSPIRE as of 10 Oct 2017



%\cite{Banks:1981nn}
\bibitem{Banks:1981nn} 
  T.~Banks and A.~Zaks,
  %``On the Phase Structure of Vector-Like Gauge Theories with Massless Fermions,''
  Nucl.\ Phys.\ B {\bf 196}, 189 (1982).
  % doi:10.1016/0550-3213(82)90035-9
  %%CITATION = doi:10.1016/0550-3213(82)90035-9;%%
  %828 citations counted in INSPIRE as of 10 Oct 2017



%\cite{Nogradi:2016qek}
\bibitem{Nogradi:2016qek} 
  D.~Nogradi and A.~Patella,
  %``Strong dynamics, composite Higgs and the conformal window,''
  Int.\ J.\ Mod.\ Phys.\ A {\bf 31}, no. 22, 1643003 (2016)
  % doi:10.1142/S0217751X1643003X
  [arXiv:1607.07638 [hep-lat]].
  %%CITATION = doi:10.1142/S0217751X1643003X;%%
  %11 citations counted in INSPIRE as of 10 Oct 2017



%\cite{Luscher:2010iy}
\bibitem{Luscher:2010iy} 
  M.~Lüscher,
  %``Properties and uses of the Wilson flow in lattice QCD,''
  JHEP {\bf 1008}, 071 (2010)
  Erratum: [JHEP {\bf 1403}, 092 (2014)]
  % doi:10.1007/JHEP08(2010)071, 10.1007/JHEP03(2014)092
  [arXiv:1006.4518 [hep-lat]].
  %%CITATION = doi:10.1007/JHEP08(2010)071, 10.1007/JHEP03(2014)092;%%
  %396 citations counted in INSPIRE as of 10 Oct 2017



%\cite{Fodor:2012td}
\bibitem{Fodor:2012td} 
  Z.~Fodor, K.~Holland, J.~Kuti, D.~Nogradi and C.~H.~Wong,
  %``The Yang-Mills gradient flow in finite volume,''
  JHEP {\bf 1211}, 007 (2012)
  % doi:10.1007/JHEP11(2012)007
  [arXiv:1208.1051 [hep-lat]].
  %%CITATION = doi:10.1007/JHEP11(2012)007;%%
  %83 citations counted in INSPIRE as of 10 Oct 2017



%\cite{Fodor:2012qh}
\bibitem{Fodor:2012qh} 
  Z.~Fodor, K.~Holland, J.~Kuti, D.~Nogradi and C.~H.~Wong,
  %``The gradient flow running coupling scheme,''
  PoS LATTICE {\bf 2012}, 050 (2012)
  [arXiv:1211.3247 [hep-lat]].
  %%CITATION = ARXIV:1211.3247;%%
  %30 citations counted in INSPIRE as of 10 Oct 2017



%\cite{Baikov:2016tgj}
\bibitem{Baikov:2016tgj} 
  P.~A.~Baikov, K.~G.~Chetyrkin and J.~H.~Kühn,
  %``Five-Loop Running of the QCD coupling constant,''
  Phys.\ Rev.\ Lett.\  {\bf 118}, no. 8, 082002 (2017)
  % doi:10.1103/PhysRevLett.118.082002
  [arXiv:1606.08659 [hep-ph]].
  %%CITATION = doi:10.1103/PhysRevLett.118.082002;%%
  %68 citations counted in INSPIRE as of 10 Oct 2017



%\cite{Herzog:2017ohr}
\bibitem{Herzog:2017ohr} 
  F.~Herzog, B.~Ruijl, T.~Ueda, J.~A.~M.~Vermaseren and A.~Vogt,
  %``The five-loop beta function of Yang-Mills theory with fermions,''
  JHEP {\bf 1702}, 090 (2017)
  % doi:10.1007/JHEP02(2017)090
  [arXiv:1701.01404 [hep-ph]].
  %%CITATION = doi:10.1007/JHEP02(2017)090;%%
  %29 citations counted in INSPIRE as of 10 Oct 2017



%\cite{Fodor:2014cxa}
\bibitem{Fodor:2014cxa} 
  D.~Nogradi, Z.~Fodor, K.~Holland, J.~Kuti et al.,
  %``The lattice gradient flow at tree level,''
  PoS LATTICE {\bf 2014}, 328 (2014)
  [arXiv:1410.8801 [hep-lat]].
  %%CITATION = ARXIV:1410.8801;%%
  %5 citations counted in INSPIRE as of 10 Oct 2017



%\cite{Fodor:2014cpa}
\bibitem{Fodor:2014cpa} 
  Z.~Fodor, K.~Holland, J.~Kuti et al.,
  %``The lattice gradient flow at tree-level and its improvement,''
  JHEP {\bf 1409}, 018 (2014)
  % doi:10.1007/JHEP09(2014)018
  [arXiv:1406.0827 [hep-lat]].
  %%CITATION = doi:10.1007/JHEP09(2014)018;%%
  %44 citations counted in INSPIRE as of 10 Oct 2017



%\cite{Hansen:2017ejh}
\bibitem{Hansen:2017ejh} 
  M.~Hansen, V.~Drach and C.~Pica,
  %``SU(3) sextet model with Wilson fermions,''
  Phys.\ Rev.\ D {\bf 96}, no. 3, 034518 (2017)
  % doi:10.1103/PhysRevD.96.034518
  [arXiv:1705.11010 [hep-lat]].
  %%CITATION = doi:10.1103/PhysRevD.96.034518;%%
  %1 citations counted in INSPIRE as of 10 Oct 2017



%\cite{Hansen:2016sxp}
\bibitem{Hansen:2016sxp} 
  M.~Hansen and C.~Pica,
  %``Sextet Model with Wilson Fermions,''
  PoS LATTICE2016 213
  [arXiv:1610.08072 [hep-lat]].
  %%CITATION = ARXIV:1610.08072;%%
  %2 citations counted in INSPIRE as of 10 Oct 2017



%\cite{Drach:2015sua}
\bibitem{Drach:2015sua} 
  V.~Drach et al.,
  %``Conformal symmetry vs. chiral symmetry breaking in the SU(3) sextet model,''
  PoS LATTICE {\bf 2015}, 223 (2016)
  [arXiv:1508.04213 [hep-lat]].
  %%CITATION = ARXIV:1508.04213;%%
  %4 citations counted in INSPIRE as of 10 Oct 2017



%\cite{Fodor:2016pls}
\bibitem{Fodor:2016pls} 
  Z.~Fodor, K.~Holland, J.~Kuti et al.,
  %``Status of a minimal composite Higgs theory,''
  PoS LATTICE {\bf 2015}, 219 (2016)
  [arXiv:1605.08750 [hep-lat]].
  %%CITATION = ARXIV:1605.08750;%%
  %9 citations counted in INSPIRE as of 10 Oct 2017



%\cite{Fodor:2016wal}
\bibitem{Fodor:2016wal} 
  Z.~Fodor, K.~Holland, J.~Kuti et al.,
  %``Electroweak interactions and dark baryons in the sextet BSM model with a composite Higgs particle,''
  Phys.\ Rev.\ D {\bf 94}, no. 1, 014503 (2016)
  % doi:10.1103/PhysRevD.94.014503
  [arXiv:1601.03302 [hep-lat]].
  %%CITATION = doi:10.1103/PhysRevD.94.014503;%%
  %16 citations counted in INSPIRE as of 10 Oct 2017



%\cite{Fodor:2015zna}
\bibitem{Fodor:2015zna} 
  Z.~Fodor, K.~Holland, J.~Kuti et al.,
  %``The running coupling of the minimal sextet composite Higgs model,''
  JHEP {\bf 1509}, 039 (2015)
  % doi:10.1007/JHEP09(2015)039
  [arXiv:1506.06599 [hep-lat]].
  %%CITATION = doi:10.1007/JHEP09(2015)039;%%
  %26 citations counted in INSPIRE as of 10 Oct 2017



%\cite{Cheng:2014jba}
\bibitem{Cheng:2014jba} 
  A.~Cheng, A.~Hasenfratz, Y.~Liu, G.~Petropoulos and D.~Schaich,
  %``Improving the continuum limit of gradient flow step scaling,''
  JHEP {\bf 1405}, 137 (2014)
  % doi:10.1007/JHEP05(2014)137
  [arXiv:1404.0984 [hep-lat]].
  %%CITATION = doi:10.1007/JHEP05(2014)137;%%
  %36 citations counted in INSPIRE as of 10 Oct 2017



%\cite{Hasenfratz:2016dou}
\bibitem{Hasenfratz:2016dou} 
  A.~Hasenfratz and D.~Schaich,
  %``Nonperturbative beta function of twelve-flavor SU(3) gauge theory,''
  arXiv:1610.10004 [hep-lat].
  %%CITATION = ARXIV:1610.10004;%%
  %9 citations counted in INSPIRE as of 10 Oct 2017



%\cite{Hasenfratz:2017mdh}
\bibitem{Hasenfratz:2017mdh} 
  A.~Hasenfratz, C.~Rebbi and O.~Witzel,
  %``Testing Fermion Universality at a Conformal Fixed Point,''
  arXiv:1708.03385 [hep-lat].
  %%CITATION = ARXIV:1708.03385;%%
  %2 citations counted in INSPIRE as of 10 Oct 2017



%\cite{Fodor:2016zil}
\bibitem{Fodor:2016zil} 
  Z.~Fodor, K.~Holland, J.~Kuti et al.,
  %``Fate of the conformal fixed point with twelve massless fermions and SU(3) gauge group,''
  Phys.\ Rev.\ D {\bf 94}, no. 9, 091501 (2016)
  % doi:10.1103/PhysRevD.94.091501
  [arXiv:1607.06121 [hep-lat]].
  %%CITATION = doi:10.1103/PhysRevD.94.091501;%%
  %18 citations counted in INSPIRE as of 10 Oct 2017



%\cite{Fodor:2017gtj}
\bibitem{Fodor:2017gtj} 
  Z.~Fodor, K.~Holland, J.~Kuti, D.~Nogradi and C.~H.~Wong,
  %``Extended investigation of the twelve-flavor $\beta$-function,''
  arXiv:1710.09262 [hep-lat].
  %%CITATION = ARXIV:1710.09262;%%

\end{thebibliography}
\end{document}